\documentclass[aps,prl,letterpaper,amsmath,amssymb,superscriptaddress,nofootinbib,twocolumn]{revtex4-2}

\usepackage{graphicx}
\usepackage{dcolumn}
\usepackage{bm}
\usepackage[usenames]{color}
\usepackage[colorlinks=true,linkcolor=blue,citecolor=blue,anchorcolor=green,urlcolor=blue]{hyperref}
\usepackage{MnSymbol}
\usepackage[percent]{overpic}
\usepackage{tabularx}
\usepackage{multirow}
\usepackage{CJK}
\usepackage{orcidlink}
\usepackage{booktabs}
\usepackage{makecell}
\usepackage{float}
\usepackage{soul}
\renewcommand{\selectlanguage}[1]{}

\def\be{\begin{equation}}
\def\ee{\end{equation}}
\def\bea{\begin{eqnarray}}
\def\eea{\end{eqnarray}}

\def\ba{\begin{aligned}}
\def\ea{\end{aligned}}

\begin{document}

\begin{CJK*}{UTF8}{}
\title{Thermally activated detection of dark particles in a weakly coupled \\quantum Ising ladder
}

\author{Yunjing Gao\orcidlink{0000-0002-1727-2577}
}
\thanks{These authors contributed equally to the work.}
\affiliation{Tsung-Dao Lee Institute, Shanghai Jiao Tong University, Shanghai 201210, China}

\author{Jiahao Yang\orcidlink{0000-0001-7670-2218}
}
\thanks{These authors contributed equally to the work.}
\affiliation{Tsung-Dao Lee Institute, Shanghai Jiao Tong University, Shanghai 201210, China}
\affiliation{International Center for Quantum Materials, School of Physics, Peking University, Beijing 100871, China}

\author{Huihang Lin\orcidlink{0000-0002-4439-6188}
}
\affiliation{Department of Physics and Beijing Key Laboratory of Opto-electronic
Functional Materials and Micro-nano Devices, Renmin University of China,
Beijing 100872, China}
\affiliation{Key Laboratory of Quantum State Construction and Manipulation (Ministry of Education), Renmin University of China, Beijing, 100872, China}

\author{Rong Yu\orcidlink{0000-0001-5936-1159}
}
\affiliation{Department of Physics and Beijing Key Laboratory of Opto-electronic
Functional Materials and Micro-nano Devices, Renmin University of China,
Beijing 100872, China}
\affiliation{Key Laboratory of Quantum State Construction and Manipulation (Ministry of Education), Renmin University of China, Beijing, 100872, China}

\author{Jianda Wu\orcidlink{0000-0002-3571-3348}
}
\email{wujd@sjtu.edu.cn}
\affiliation{Tsung-Dao Lee Institute, Shanghai Jiao Tong University, Shanghai 201210, China}
\affiliation{Shanghai Branch, Hefei National Laboratory, Shanghai 201315, China}
\affiliation{School of Physics and Astronomy, Shanghai Jiao Tong University, Shanghai 200240, China}

\begin{abstract}
The Ising$_h^2$ integrable field theory emerges when two quantum
critical Ising chains are weakly coupled.
This theory
possesses eight types of relativistic particles,
among which the lightest one ($B_1$)
has been predicted to be a dark particle,
which cannot be excited from the ground state through
(quasi-)local operations.
The stability on one hand
highlights its potential for applications,
and on the other hand makes it challenging to be observed.
Here, we point out that the mass of the $B_1$ dark particle $m_{B_1}$ appears
as a thermally activated gap extracted
from local spin dynamical structure factor at
low frequency ($\omega \ll m_{B_1}$) and low temperatures
($T \ll m_{B_1}$). We then
further propose that this gapped behavior can be directly
detected via
the NMR relaxation rate measurement
in a proper experimental setup.
Our results provide a practical criterion for
verifying the existence of dark particles.
\end{abstract}

\date{\today}
\maketitle
\end{CJK*}

\paragraph*{Introduction.---}
Quantum criticality captures collective behaviors of many-body systems within the quantum critical region,
regardless of their microscopic origins \cite{sachdev_2011,Altland}.
At certain quantum critical point,
conformal invariance may emerge associated
with special scaling for dynamical and thermal behaviors
\cite{Polyakov,BELAVIN1984333,Friedan}.
Relevant perturbations at this point may
result in new quantum integrability
with exactly solvable spectrum and
many-body wavefunctions
\cite{ZamE8, coupleCFT, Smirnov, Mussardobook}.
A paradigmatic model is
the transverse field Ising chain (TFIC),
whose quantum critical point (QCP) is governed by
a central charge $1/2$ conformal field theory \cite{BELAVIN1984333}.
Introducing Ising field perturbation into quantum critical TFIC further
leads to the emergence of quantum $E_8$ integrable field theory
\cite{ZamE8,JD2014}.
Recently,
these theoretical predictions have been confirmed in
quasi-1D magnetic materials through combined efforts from experiments \cite{PhysRevLett.120.207205,PhysRevB.101.220411,PhysRevX.4.031008,E8,WANG20242974,Yang_2023}.

Quantum integrable systems can also be categorized within the framework of coupled conformal field theory.
Specifically,
for the quantum Ising ladder composed of two weakly-coupled critical TFICs,
its low-energy physics is described by the Ising$_h^2$ integrable field theory (IIFT) \cite{coupleCFT}.
Among the 8 types of relativistic particles possessed by the IIFT,
three of them (labeled as $B_{1,3,5}$) are 
``dark particles'' \cite{dark},
which are inherently prohibited from the ground state 
through any local or quasi-local spin operations, due to global selection rules.
Remarkably,
the lightest dark particle is theorized to be robust and long lifetime once prepared, thus may have
potential applications in quantum information and quantum technology.
The absence of single dark particle peaks
in the zero-temperature
dynamical structure factor (DSF) has been
validated numerically~\cite{dark,Xning}.
An additional advantage lies in the model construction,
which can be simulated in Rydberg arrays or realized in
quasi-1D magnets, e.g., CoNb$_2$O$_6$.
However,
although dark particles contribute to the spectra through multi-particle channels,
their spectral signatures are expected to remain elusive in spectroscopy 
of materials predicted to exhibit Ising$_h^2$ physics \cite{Xning}.
Given the intriguing properties and significant potential applications of dark particles,
it is desired to have a proper experimental
setup capable of directly confirming the presence of single dark particles.

In this letter,
after introducing the IIFT and emergent dark particles,
we propose nuclear magnetic resonance (NMR) relaxation rate measurement
at low temperatures, where the lightest
dark particle's mass
manifests as the thermal activated gap extracted from
the measurement.
Quantitatively,
we analyze asymptotic behaviors of the local spin DSF in low-frequency and low-temperature limit.
To connect field theory result and lattice simulation,
we analytically determine the relation between the gap and the interchain coupling.
Our results show that the thermal activation gap extracted from the local DSF directly corresponds to the mass of the lightest dark particle $m_{B_1}$,
instead of the lightest visible particle $m_{B_2} \approx 1.95 m_{B_1}$
measured in other dynamical spectroscopies \cite{Morris2021b,ZWang,Coldea}.
The characteristic feature can serve as a distinctive sign to confirm the existence of the lightest dark particle $B_1$.

\paragraph*{The model.---}
\label{secII:model}
\begin{figure}[htp]
    \centering
    \includegraphics[width=0.45\textwidth]{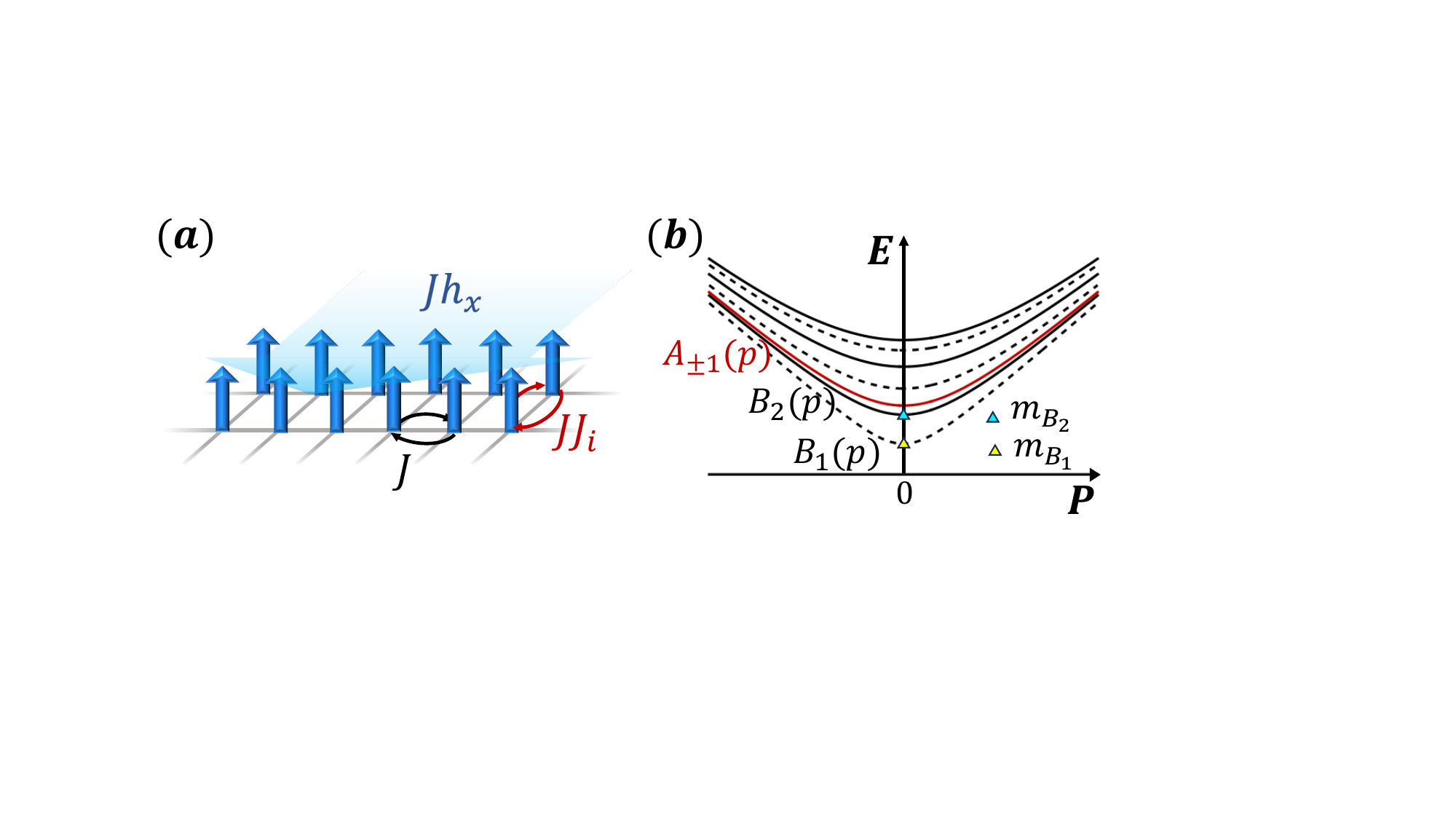}
    \caption{(a) Illustration of two weakly coupled Ising chains
    with transverse field $Jh_x$. (b) Single particle dispersion of the IIFT. The red line denotes $A_{\pm1}$. The dotted black lines denote $B_1$, $B_3$, $B_5$ and the black lines denote $B_2$, $B_4$ and $B_6$ from below to above, respectively. The intersection of the dispersions with the $P=0$ line reveals the particle mass.}
    \label{fig:ill}
\end{figure}

The Hamiltonian of two weakly-coupled quantum critical TFICs [Fig.~\ref{fig:ill}(a)]follows
\be
H=
-J\left(\sum_{l=1,2}\sum_{n=1}^{N-1}
(\sigma_{n}^{z(l)}\sigma_{n+1}^{z(l)}
+ h_x \sigma_{n}^{x(l)})
+\sum_{n=1}^N
J_i\sigma_{n}^{z(1)}\sigma_{n}^{z(2)}\right),
\label{eq:Ham}
\ee
where $\sigma_{n}^{\mu(1,2)}$ are Pauli matrices
associated with spin operators $S^\mu = \sigma^\mu/2\  (\hbar = 1, {\mu=x,y,z})$ at site
$n$ on chain (1) or (2).
$J$ and $J_i$ are intra- and inter-chain couplings, respectively.
Here we focus on $h_x=1$ which is the critical field for each decoupled chain.

For $J_i=0$,
Jordan-Wigner transformation maps spins from each chain into a set of fermions, which can be
further diagonalized by the Bogoliubov transformation \cite{pfeuty_one-dimensional_1970, Lieb1961}.
In the scaling limit,
where the lattice spacing $a\to0$ and $J\to\infty$ with $2Ja=1$,
the discrete fermions are recast as free fermionic field,
leading to the central charge 1/2 conformal field theory $H_{c=1/2}$ \cite{QFTising,2isingboson}.
In parallel,
scaling limit of $\sigma^{z}_j$ and $\sigma^{x}_j$ can be taken,
conventionally referred to as order operator $\sigma(x)$ and energy operator $\epsilon(x)$.
Historically,
Bosonization is performed by combining the two sets of real fermions from each chain,
leading to a free model for the bosonic field $\phi$ \cite{2isingboson}.
Operator correspondences that will be useful are listed as
$\sigma^{(1)}(x)\sigma^{(2)}(x)\rightarrow\,:\cos\phi(x)/2:$,
$\epsilon^{(1)}(x)+\epsilon^{(2)}(x)\rightarrow\, :\cos\phi(x):$ and
$\epsilon^{(1)}(x)-\epsilon^{(2)}(x)\rightarrow\, :\cos\Theta(x):$,
where the dual field $\Theta(x)$ satisfies
$\partial\Theta(x)/\partial x=-\partial \phi(x)/\partial t$.

In the presence of weak interchain coupling,
the perturbed Hamiltonian follows
\be\ba
H = H_{c=1/2}^{(1)}+H_{c=1/2}^{(2)}+\gamma\int dx \sigma^{(1)}(x)\sigma^{(2)}(x),
\label{eq:pertbCFT}
\ea\ee
which gives the Ising$_h^2$ integrable field theory (IIFT),
with $(1,2)$ referring to chain labels and $\gamma$ is the rescaled interchain coupling.
Its quantum integrability is more transparent
by mapping to the bosonization form \cite{coupleCFT}
\be
H_{\text{B}}=\int dx \left[\frac{1}{2}\left(\frac{\partial\phi}{\partial t}\right)^2+\frac{1}{2}\left(\frac{\partial\phi}{\partial x}\right)^2+\lambda\cos\frac{\phi(x)}{2}\right],
\label{eq:isingh2}
\ee
with $\lambda = \sqrt 2 \gamma$ \cite{BASEILHAC2001607}.
Note that Eq.~\eqref{eq:isingh2} takes the same form as the sine-Gordon model \cite{coleman},
with $\phi$ field in the former case defined on a $\mathbb{Z}_2$ orbifold.
More details can be found in Ref.~\cite{SM}.

\paragraph{The $D_8^{(1)}$ particles.---}
Spectrum and scattering matrices can be exactly solved for Eq.~\eqref{eq:isingh2},
characterized by the $D_8^{(1)}$ algebra \cite{coupleCFT}.
The IIFT theory \cite{coupleCFT} accommodates 8 types of particles: 6 breathers
($B_n$, $n=1,2,\cdots,6$), a soliton ($A_{+1}$) and an anti-soliton ($A_{-1}$).
Their masses are related as $m_{B_n}=m_{B_1}\sin(n\pi/14)/\sin(\pi/14)$
and $m_{A_{\pm}}=m_{B_1}/(2\sin(\pi/14))$.
Each particle follows relativistic dispersion relation, i.e.,
$E=\sqrt{m^2+p^2}$,
where the ``speed of light'' is set as 1.
In the context of quantum integrable field \cite{Smirnov,Mussardobook},
all excitations within the model can be described by single particles or their combinations,
referred to as asymptotic states.
An $r$-particle state is denoted as $|F_1(\theta_1)\dots F_r(\theta_r)\rangle$,
with particle type $F$ and rapidity $\theta$,
which carries eigenenergy $E_{\{r\}}=\sum_{j=1}^r m_{P_j}\cosh\theta_j$ and momentum
$P_{\{r\}}=\sum_{j=1}^r m_{P_j}\sinh\theta_j$.
Scatterings between different states can be factorized
into two particle cases which are known in \cite{coupleCFT},
setting up a systematical way to determine transition matrices through local field operators.

Global properties in the IIFT restrict the excitation channels.
Taking advantage of the sine-Gordon model \cite{bookBosonization},
topological charge $Q$ can be assigned to the soliton and antisoliton as $+1$ and $-1$, respectively,
while $Q=0$ for the breathers.
And for the parity conjugation $\mathcal{C}$ that operates as $\mathcal{C}\phi\mathcal{C}^{-1}=-\phi$,
we have
$\mathcal{C}|B_{n}(\theta)\rangle=(-1)^n|B_{n}(\theta)\rangle$
and $\mathcal{C}|A_{\pm 1}(\theta)\rangle=|A_{\mp1}(\theta)\rangle$.
As studied in Ref.~\cite{dark},
excitation
from the even-parity ground state
to the odd-parity $B_{1,3,5}$ via any local
or quasi-local spin operations is forbidden.
The $B_{1,3,5}$ are referred to as dark particles.
On the other hand,
the $\sin\phi(x)$ operator which connects
odd-parity states with the ground state
corresponds to $\prod_{j=1}^l\sigma_j^{z(1)}\sigma_j^{z(2)}$ $(x=la)$ in the spin representation.
Since it involves macroscopic number of spins,
such operation is strongly suppressed in the vacuum fluctuations
since the coherent length of the fluctuation is exponentially cutoff
because of the gap.
As a result,
the lightest particle $B_1$ cannot spontaneously decay once being prepared.

\paragraph*{Thermal activated measurement.---}
Spin DSF measurements at low temperatures ($T \ll m_{B_1}$),
such as inelastic neutron scattering (INS) \cite{E8,WANG20242974} and THz spectroscopy \cite{ZWang},
are dominant by excitations directly from the ground state.
Following discussion in the previous paragraph,
energy levels of the dark particles cannot be
resolved in these measurements.
In contrast,
we propose that the thermal activation behavior observed in NMR measurement can directly probe the lightest dark particle.
Different from aforementioned detections that cover a rather larger energy range (e.g., $0< \omega \lesssim 6 m_{B_1} $ in Ref.~\cite{ZWang}),
the NMR has low-energy detectability and measures locally.
With input energy much smaller than $m_{B_2}$,
excitation from the ground state cannot happen,
leading to dominant excitation from $B_1(0)$ to $B_1(p)$ [Fig.~\ref{fig:ill}(b)].
Intensity of such excitation is affected by the thermal distribution of $B_1$,
reflecting in the thermal activation gap that can be determined experimentally in a relaxation process.

To capture the thermal activation gap,
first we introduce the local spin DSF at finite temperature as following
\be
C^{\mathcal{O}}(\omega,T) = \int_{-\infty}^{\infty}dt \,e^{i\omega t}
\langle\mathcal{O}(t,0)\mathcal{O}^{\dagger}(0,0)\rangle_T,
\label{eq:ODSF}
\ee
where $\mathcal{O}$ denotes the field theory counterpart of local spin operators.
$t$ in the integral denotes real time,
$\omega$ and $T$ denote the frequency and temperature, respectively.
In the NMR measurement,
the spin-lattice relaxation rate is given by \cite{10.1143/PTP.28.371,PhysRevB.106.125149}
\be
\frac{1}{T_1}\sim|A_y|^2C^{y}(\omega_n)+|A_z|^2C^{z}(\omega_n),
\ee
with hyperfine coupling constant $A_j(j=x,y,z)$ and resonant frequency $\omega_n$ of NMR measurement.
On the other hand,
$C^{x}$ is related to the spin-spin relaxation rate
$1/T_2 = 1/T_1^{\prime}+1/T_2^{\prime}$,
where
\be
\frac{1}{T_1^{\prime}}=\frac{A}{T_1},\quad
\frac{1}{T_2^{\prime}} = |A_x|^2C^{x} (\omega\to 0),
\ee
with constant $A$ depending on microscopic details.
In the next paragraph,
we determine the thermal behavior of the local spin DSF.

\paragraph*{Dark particle detection.---}
\label{secIII:DSF}

\begin{figure}[b]
    \centering
    \includegraphics[width=0.45\textwidth]{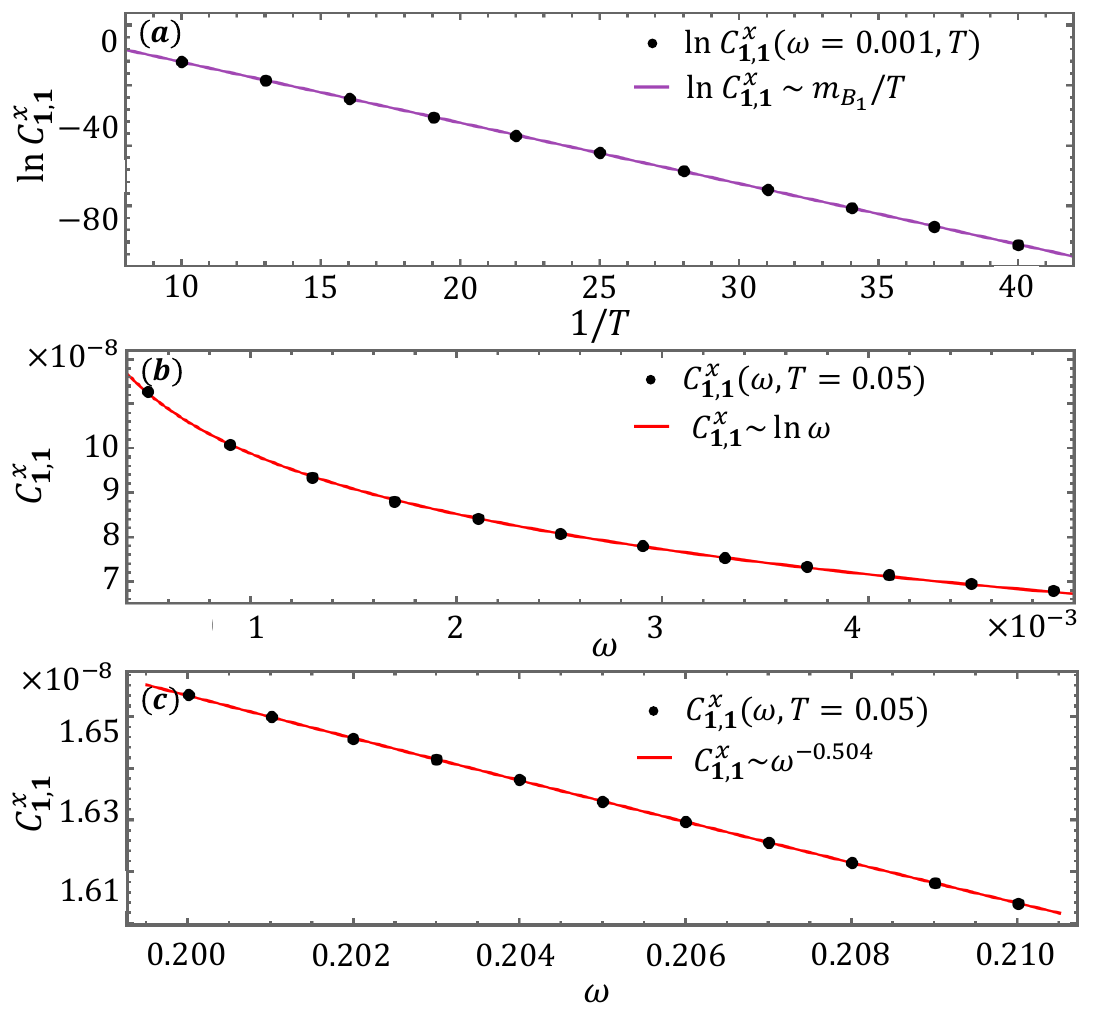}
    \caption{Asymptotic behaviors of $C^x_{1,1}(\omega,T)$ [Eq.~\eqref{eq: S11}] calculated from numerical integration with $m_{B_1}=1$. (a) shows the result of fixed energy $\omega$ and the fitted curve $\ln C^x_{1,1}=-1.009/T+4.083$. (b) and (c) show the results with fixed temperature $T$ in the region $\omega\ll T$ and $\omega\gg T$, respectively. The curve in (b) is fitted as $C_{1,1}^x\times10^{8}=-1.960\ln\omega+3.647$, and in (c) $C_{1,1}^x\times10^{8}=0.736\omega^{-0.504}$.}
    \label{fig:2}
\end{figure}
We first study the
transverse spin DSF $C^x$ for finite but small energy ($\omega\ll m_{B_1})$ at low temperature.
According to previous paragraph,
$\sigma^x$ corresponds to $\cos\phi+\cos\Theta$ in the bosonization.
The operation of $\cos\phi$ preserves topological charge while
$\cos\Theta$ connects states with topological charge-1 difference.
Terms with maximum Boltzmann weight resulted from ground state excitations
only contribute to the DSF after reaching $\omega_\text{threshold} = m_{B_2} \approx1.95m_{B_1}$,
with channel $\langle 0|\cos\phi|B_2(0)\rangle$,
which vanishes as energy conservation can not hold for $\omega \ll m_{B_2}$.
Through the cluster expansion \cite{Pozsgay_2010},
the leading term appears at the order of $e^{-m_{B_1}/T}$
corresponding to $|B_1\rangle\to|B_1\rangle$ channel \cite{SM},
\be\ba
&C^{x}_{1,1}(\omega,T)\approx\sumint_{s=\pm 1}\frac{d\theta}{2\pi}
\frac{e^{-\frac{m_{B_1}}{T}\cosh\theta}|\langle B_1(\theta)|\cos\phi|B_1(s\tilde{\theta})\rangle|^2}{\sqrt{(\omega+m_{B_1}\cosh\theta)^2-m_{B_1}^2}}\\
\label{eq: S11}
\ea\ee
with $\cosh\tilde{\theta}=(\omega+m_{B_1}\cosh\theta)/m_{B_1}$.
The leading behaviors of the integral in Eq.~\eqref{eq: S11} can be obtained analytically \cite{PhysRevLett.113.247201}.
For $\omega\ll T\ll m_{B_1}$,
\be
C^x_{1,1}(\omega,T)\approx\frac{e^{\omega/2T}|F^{11}|^2}{\pi m_{B_1}}e^{-m_{B_1}/T}\left(-\ln\frac{\omega}{4T}-\gamma_E\right),
\label{eq:C_11}
\ee
and for $T\ll \omega\ll m_{B_1}$,
\be
C^x_{1,1}(\omega,T)\approx\frac{|F^{11}|^2}{\pi m_{B_1}}e^{-m_{B_1}/T}\left[\sqrt{\frac{\pi T}{\omega}}-\frac{\sqrt{\pi}}{4}\left(\frac{T}{\omega}\right)^{\frac{3}{2}}\right],
\ee
where $F^{11}=|\langle 0|\cos\phi|B_1(i\pi)B_1(0)\rangle|^2=31.756$,
originating in the transition element $\langle B_1(0)|\cos\phi|B_1(0)\rangle$.
In both limits,
$e^{-m_{B_1}/T}$ behavior is obtained for fixed $\omega$ [Fig.~\ref{fig:2} (a)],
resulting in the observable thermal activation gap $m_{B_1}$.
For the isothermal case,
logarithmic divergence in $\omega$ is found for $\omega\ll T$ [Fig.~\ref{fig:2} (b)] in contrast to the power law behavior for $\omega\gg T$ [Fig.~\ref{fig:2} (c)].
Moreover,
the subleading contribution comes from the transition $|B_2\rangle\to|B_2\rangle$,
which is negligible since the corresponding Boltzmann weight $e^{-m_{B_2}/T}\approx e^{-1.95m_{B_1}/T}
\ll e^{-m_{B_1}/T}$
for $T \ll m_{B_1}$.

\begin{figure}
    \centering
    \includegraphics[width=0.42\textwidth]{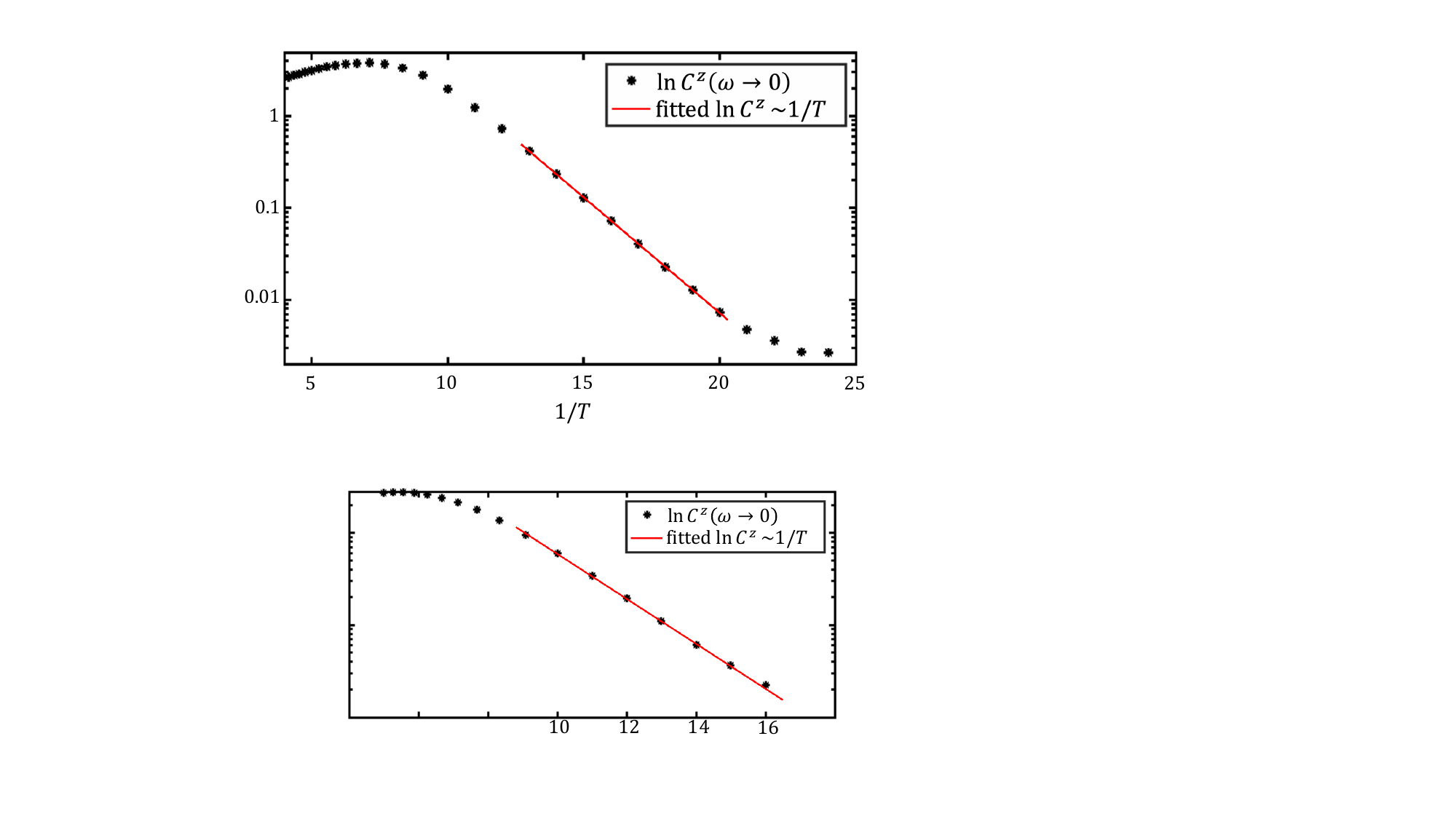}
    \caption{
    Temperature dependence of $C^z(\omega\to0)$ [Eq.~\eqref{eq:ODSF}] from QMC simulation with $J_i=0.1$, $J=1$ on a $2\times4096$ lattice.
    Red line shows the fitted relation $C^z\sim e^{-0.58/T}$.}
    \label{fig:3}
\end{figure}

$C^z$ and $C^y$ [Eq.~\eqref{eq:ODSF}] are beyond
IIFT analytical form factor scheme,
as $\sigma^{z}$ and $\sigma^y$ do not correspond to local fields in the bosonic theory.
However, their thermal behaviors can be determined numerically.
The first symmetry-allowed channel in $C^z$ comes from $|B_1\rangle\to|B_1\rangle$,
dominated by the zero momentum mode,
which is indeed non-vanishing as elaborated in \cite{SM}
following density-matrix renormalization group (DMRG) calculation.
On the other hand,
$C^z(\omega\to 0)$ can be directly obtained by quantum Monte Carlo (QMC) simulation \cite{newmanb99},
which is calculated from $2(\pi \tau)^{-1}\sum_j\langle\delta
S_i^z((2\tau)^{-1}) \delta S_i^z(0)\rangle$.
Here $\tau$ is the imaginary time
and $\delta S_i^z =S_i^z - \langle S_i^z \rangle$ \cite{prr,QMC}.
The thermal activation gap fitted from Fig.~\ref{fig:3}
is about $0.58J$ at $J_i=0.1$.
However, this number can not
compare directly with $m_{B_1}$ in the IIFT.
For achieving the goal
we need to analytically determine $m_{B_1}$ in the
corresponding lattice model.

\begin{figure}[b]
    \centering
    \includegraphics[width=0.42\textwidth]{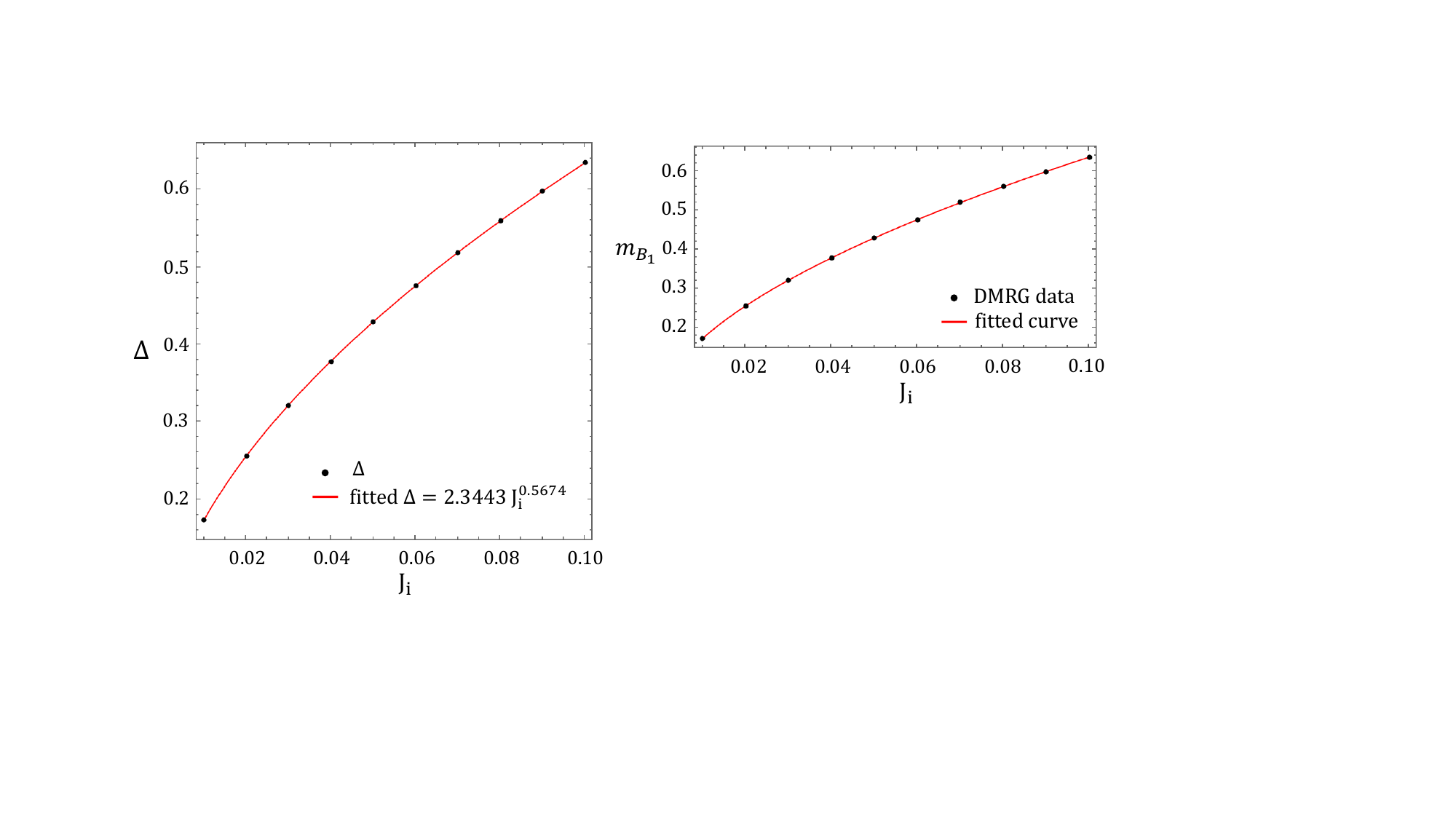}
    \caption{Relation between $m_{B_1}$ and $J_i$. The black dots are calculated from DMRG method with $N=400$ for each chain and $J=1$ with fitted result $m_{B_1}=2.3443J_i^{0.5674}$ (red line). It agrees well with analytical result $m_{B_1}=2.3797J_i^{4/7}$ from Eq.~\eqref{eq:gapji}.}
    \label{fig:scale}
\end{figure}
The IIFT predicts the mass of $B_1$ as $m_{B_1}=(\lambda/\mathcal{C}_1)^{4/7}$,
with $\mathcal{C}_1=0.33645\,\cdots$ \cite{BASEILHAC2001607}.
Following similar strategy as in \cite{PhysRevB.110.195101},
we fix the normalization condition for the lattice model [Eq.~\eqref{eq:Ham}]
and its scaling limit [Eq.~\eqref{eq:pertbCFT}], i.e.,
$\left\langle\gamma\int dx\sigma^{(1)}(x)\sigma^{(2)}(x)\right\rangle=\left\langle J\sum_{n=1}^NJ_i\sigma_n^{z(1)}\sigma_n^{z(2)}\right\rangle$,
which is recovered as $\gamma L\langle\sigma\rangle^2=
J J_iN\langle\sigma^{z}\rangle^2$ in the decoupled limit.
Relation between $\gamma$ and $J_i$ can be obtained by
inserting the results for magnetizations ($\langle\sigma^z\rangle$ and $\langle\sigma\rangle$) \cite{McCoyWu1973}.
Then we arrive at
\be
m_{B_1} = \left(\frac{2\sqrt{2}}{\mathcal{C}_12^{1/6}e^{-1/4}\mathcal{A}^{3}}\right)^{4/7}J_i^{4/7} J,
\label{eq:gapji}
\ee
where the Glaisher's constant $\mathcal{A}=1.282427\cdots$.
More details can be found in \cite{SM}.
Coefficients determined here are consistent with numerical calculation [Fig.~\ref{fig:scale}],
where $m_{B_1}$ is obtained from the energy of the first excited state,
following the DMRG method \cite{SCHOLLWOCK,PhysRevLett.69.2863}.

Now we observe that the fitted gap from QMC
is close to the predicted $m_{B_1}=0.6384J$ obtained from Eq.~\eqref{eq:gapji}.
The small deviation comes from finite temperature and finite size effect.
Furthermore,
the general relation for our model $C^y=(\omega/J)^2 C^z/4$ \cite{JD2014, dark} implies that
$\sigma^y$ channel is negligible when $\omega \ll J$.
Consequently, extracting the thermal activation gap from experiments allows for
the verification of the existence of $B_1$.

\paragraph*{Discussions.---}
Proper Ising-chain compounds may serve as to study dark particles.
For instance,
the quasi-1D magnet CoNb$_2$O$_6$,
which was claimed to accommodate $E_8$ physics \cite{Coldea, ZWang}.
The $E_8$ model emerges in a quantum critical Ising chain under the perturbation of a longitudinal field along the Ising spin direction \cite{ZamE8}.
In their setup,
the transverse field is tuned to the putative 1D TFIC QCP in the 3D ordering dome at low temperature ($T \ll T_N$),
where the 3D order is considered to effectively provide the longitudinal field coupled to the transverse-field Ising chain in the material.
Recent analysis \cite{Xning} shows that
the physics here is more comprehensively described by a quantum Ising ladder with $\mathcal{D}_8^{(1)}$ structure.
As the putative 1D QCP lives in the vicinity of the 3D QCP in the material,
a 1D model is not sufficient for treating 3D fluctuations.
Moreover,
the aforementioned effective field from 3D ordering is suppressed by the strong magnetic frustration in the material.

For the model described in Eq.~\eqref{eq:Ham},
the applied transverse field can serve as static magnetic field in an NMR setup \cite{PhysRevX.4.031008}.
A proper NMR measurement will not only reveal the existence of the lightest dark particle $B_1$
in the CoNb$_2$O$_6$,
but also provide evidence to confirm the $\mathcal{D}_8^{(1)}$ physics of the material,
as the $E_8$ theory has no such hidden energy levels.
Apart from the magnetic materials,
it is also possible to directly simulate
the Hamiltonian in Eq.~\eqref{eq:Ham} in Rydberg arrays.

As a starting point for the proposed detection,
we assume that thermal equilibrium has been reached in this system.
Considering that the phonon-spin coupling is typically weak,
the perturbation would not change the ``dark'' properties.
Though direct excitation from ground state to dark particles is forbidden,
the transition from,
e.g.,
$|A_{\pm}
\rangle$ to $|B_1\rangle$ is permitted and $|A_{\pm}
\rangle$ can be excited from ground state.
In contact with thermal reservoir,
$B_1$ can be reached through a secondary process $|0\rangle\to |A_{\pm}\rangle\to |B_1\rangle$.
It is worth further investigating on the detailed thermalization procedure.

\paragraph*{Conclusions.---}
To conclude, following the form factor approach and cluster expansion in the region $\omega, T\ll m_{B_1}$,
we analytically determine the thermal activation gap for $\sigma^x$ channel,
corresponding to the mass of the lightest dark particle.
The same thermal activation gap is also
obtained for $\sigma^z$ following DMRG and QMC numerical calculations,
which is in contrast to the first peak obtained from zero temperature spin DSF.
Taking the advantage,
we propose that a proper NMR experiment can detect the lightest dark particle
through the relaxation rates $1/T_1$ and $1/T_2$
measurements where the mass of the lightest dark particle can be
extracted as thermal activation gap.
Potential material candidates, such as CoNb$_2$O$_6$ and compounds
effectively described by the Ising ladder
are suggested.
Rydberg array and STM experiments could also directly simulate the
required Ising ladder to probe the dark particles.

\section{Acknowledgments}
The work at Shanghai Jiao Tong University is supported by
the National Natural Science Foundation of China Grant Nos. 12450004, 12274288 and
the Innovation Program for Quantum Science and Technology Grant No. 2021ZD0301900.
The work at Renmin University of China is supported by the National Key R\&D Program of China (Grant No. 2023YFA1406500), and the National Natural Science Foundation of China (Grant Nos. 12334008 and 12174441).
H. L. also acknowledges the supported by the Postdoctoral Fellowship Program of CPSF (No.~GZC20241724).

\appendix
\newpage
\onecolumngrid
\setcounter{figure}{0}
\makeatletter
\renewcommand{\thefigure}{S\@arabic\c@figure}
\setcounter{equation}{0} \makeatletter
\renewcommand \theequation{S\@arabic\c@equation}
\section*{{Supplementary Material --- Thermally activated detection of dark particles in a weakly coupled quantum Ising ladder}}

\section{From the lattice model to the Ising$_h^2$ integrable field theory (IIFT)}
Consider the quantum critical transverse field Ising chain (TFIC) (Eq.(1) in the main text).
Following the Jordan-Wigner transformation,
spin operators can be mapped to fermionic operators $c_j^{\dagger}, c_j$ as \cite{Pfeuty}
\be
\sigma^z_j=(c_j^{\dagger}+c_j)e^{\pm i\pi\sum_{l<j}c_l^{\dagger}c_l},\quad\sigma^x_j=2c_j^{\dagger}c_j-1.
\label{eq:JW}
\ee
with $\{c_i,c_j\}=\delta_{ij}$ and $\{c_i,c_j\}=\{c_i^{\dagger},c_j^{\dagger}\}=0$.
Consequently,
the disorder operator is mapped as $\mu_j=\prod_{k<j}\sigma^x_k=e^{\pm i\pi\sum_{k=1}^{j-1}c_k^{\dagger}c_k}$. And we have $\sigma_j^z=(c_j^{\dagger}+c_j)\mu_j$.

Furthermore,
the Majorana spinor $\psi = (\psi_L,\psi_R)^T$ is introduced with 
left ($\psi_L$)- and right ($\psi_R$)- components \cite{2isingboson}
\be
\psi_L(j)=\frac{(-1)^j}{\sqrt{a}}(c_j^{\dagger}e^{-i\pi/4}+c_je^{i\pi/4}),\quad
\psi_R(j)=\frac{(-1)^j}{\sqrt{a}}(c_j^{\dagger}e^{i\pi/4}+c_je^{-i\pi/4}),
\label{eq:Majorana}
\ee
where $a$ is the lattice spacing.
Commutation relations follow
$\{\psi_R(i),\psi_R(j)\}=\{\psi_L(i),\psi_L(j)\}=\delta_{ij}/a$, $\{\psi_R(i),\psi_L(i)\}=0$.
Then the relations between spin operators and Majorana spinor are directly obtained as
\be
\sigma^x_j=-ai\psi_R(j)\psi_L(j),\quad
\sigma^z_j = (-1)^j\sqrt{\frac{a}{2}}\left(\psi_R(j)+\psi_L(j)\right)e^{\pm i\pi\sum_{l<j}(-ai\psi_R(l)\psi_L(l)+1)/2}.
\label{eq:sigmazpsi}
\ee
By taking the scaling limit that $a\to0$ and $Ja$ keeps finite,
the critical TFIC can be described by a free Hamiltonian in continuous space
 in terms of the Majorana spinor
\be
\mathcal{H}_{ms}=\psi^{\dagger}(x)\left(-i\gamma^5\frac{\partial}{\partial x}\psi(x)\right)
\ee
with $x=ja$, $\gamma^5=\sigma^z$ and $2Ja=1$.
In the field theory language,
$\sigma^z_j$, $\mu_j$ and $\sigma^x_j$ are referred to as order operator $\sigma(x)$, disorder operator $\mu(x)$ and energy operator $\epsilon(x)$, respectively.

Then we consider two copies of the quantum critical TFIC.
Since the Majorana spinor is real,
two sets of Majorana spinors $\psi^{(1,2)}$ obtained from the two 
chains can be further gathered into complex Dirac spiniors $\chi$ as \cite{2isingboson}
\be
\chi = \frac{1}{\sqrt{2}}\left(\psi^{(1)}+i\psi^{(2)}\right),\quad
\chi^{\dagger} = \frac{1}{\sqrt{2}}\left(\psi^{(1)}-i\psi^{(2)}\right).
\ee
Then the Dirac spinors can be bosonized into free massless bosonic field,
following the bosonization rule
\be
\chi_R = \frac{\alpha_R}{\sqrt{N}}:e^{i\phi_R}:,\quad
\chi_L = \frac{\alpha_L}{\sqrt{N}}:e^{-i\phi_L}:,
\label{eq:rule}
\ee
where the normal ordering $::$ enforces annihilation operators to the right and
$\alpha_{L,R}$ ensure the anti-commutation relation of $\chi_{R,L}$ \cite{2isingboson}.
Right- and left- going components of the bosonic field are introduced via the mode expansion
$\phi(x,t)=\phi_R(x-t)+\phi_L(x+t)$,
i.e.,
\be\ba
&\phi_R(t-x)=\frac{\phi_{0R}}{2\pi}-\frac{Q_R}{2N}(t-x)-\frac{i}{2\pi}\sum_{n\neq 0}\frac{\overline{a}_n}{n}e^{-2\pi in[(t-x)/N]}\\
&\phi_L(t+x)=\frac{\phi_{0L}}{2\pi}+\frac{Q_L}{2N}(t+x)-\frac{i}{2\pi}\sum_{n\neq 0}\frac{a_n}{n}e^{-2\pi in[(t+x)/N]},
\ea\ee
where $N$ is the system size,
and $Q_{R,L}$ are conjugation of the zero modes $\phi_{0R,L}$,
satisfying $[Q_R,\phi_{0R}]=-[Q_L,\phi_{0L}]=i/2$.
$a_{n}, \overline{a}_n$ are related to boson creation and annihilation operators ($\tilde{a}_n$ and $\tilde{a}_n^{\dagger}$) by
\be
a_n=\left\{
    \begin{aligned}
    -i\sqrt{n}\tilde{a}_n & (n>0)\\
    i\sqrt{-n}\tilde{a}_{-n}^{\dagger} & (n<0)\\
    \end{aligned}
    \right.
,\quad
\overline{a}_n=\left\{
    \begin{aligned}
    -i\sqrt{n}\tilde{a}_{-n} & (n>0)\\
    i\sqrt{-n}\tilde{a}_n^{\dagger} & (n<0)\\
    \end{aligned}
    \right.
,
   \ee
with non-vanishing commutators $[\overline{a}_n,\overline{a}_m]=[a_n,a_m]=n\delta_{n+m,0}$.
The dual field of $\phi$ is introduced as $\Theta(x,t)=\phi_R(x-t)-\phi_L(x+t)$
which satisfies
\be
\frac{\partial \Theta}{\partial x}=-\frac{\partial \phi}{\partial t}.
\label{eq:dual}
\ee
Thus we obtain the effective free bosonic field theory with the Hamiltonian density
\be
\mathcal{H}_{fb}=\partial^2 \phi(x)/\partial x^2,
\label{eq:boson}
\ee
corresponding to a conformal field theory with central charge $c = 1$ \cite{Ashkinteller}.

Taking the scaling limit of Eq.~\eqref{eq:sigmazpsi},
we have $\epsilon(x)^{(1,2)}=i\psi_R^{(1,2)}(x)\psi_L^{(1,2)}(x)$.
Several operator correspondences in the bosonization representation can be derived by
inserting the bosonization rules Eq.~\eqref{eq:rule} and using the Baker-Hausdorff formula for normal ordering operators \cite{Fradkin}
\be
:e^A::e^B:=e^{[A^+,B^-]}:e^{A+B}: \quad\text{if } [A^+,B^-]\text{ is c-number,}
\label{eq:BA}
\ee
where $\hat{O}=\hat{O}_++\hat{O}_-$ and $+/-$ denotes the creation/annihilation piece of the operator,
which are summarized in TABLE.~\ref{Tab:I} \cite{KADANOFF,2isingboson}.
\begin{table}[h]
    \setlength\tabcolsep{8pt}
    \renewcommand{\arraystretch}{1.25}
   \centering
    \begin{tabular}{ll}
    \specialrule{0em}{6pt}{6pt}
        \toprule [0.7pt]
        spin field & bosonized\\
        \midrule [0.4pt]
        $\sigma^{(1)}(x)\sigma^{(2)}(x)$ & $:\cos[\phi(x)/2]:$ \\
        $\epsilon^{(1)}(x)+\epsilon^{(2)}(x)$ & $:\cos[\phi(x)]:$\\
        $\epsilon^{(1)}(x)-\epsilon^{(2)}(x)$ & $:\cos[\Theta(x)]:$\\
        $\epsilon^{(1)}(x)\epsilon^{(2)}(x)$ & $\partial_\gamma\phi\partial^\gamma\phi \,(\gamma=x,\,t)$\\
        $\sigma^{(1)}(x)\mu^{(2)}(x)$ & $:\cos[\Theta(x)/2]:$\\
        $\mu^{(1)}(x)\mu^{(2)}(x)$ & $:\sin[\phi(x)/2]:$\\
        \bottomrule [0.7pt]
    \end{tabular}
    \caption{Bosonization correspondences for spin operators in the scaling limit.}
    \label{Tab:I}
\end{table}

Now we turn on weak interchain coupling,
the continuum theory of Eq.(1) is given by the action
\be
\mathcal{A}_{\text{Ising}_h^2}=\mathcal{A}_{c=1/2}^{(1)}+\mathcal{A}_{c=1/2}^{(2)}+\lambda^{\prime}\int dxdt\sigma^{(1)}\sigma^{(2)},
\label{eq:isingh2action}
\ee
where each chain is described by a conformal field theory $\mathcal{A}_{c=1/2}^{(1,2)}$,
and the rescaled coupling strength $\lambda^{\prime}\propto\lambda^{7/4}$.
The last term in  Eq.~\eqref{eq:isingh2action} implies that additional interaction $\cos(\phi/2)$ should be added to the free boson theory Eq.~\eqref{eq:boson},
which formally gives the Ising$_h^2$ action [Eq.~(3)].
Eq.~(3) appears in the same from as a sine-Gordon model,
while the difference lies in that the $\phi$ field is defined on a $\mathbb{Z}_2$ orbifold \cite{coupleCFT},
rather than on a circle.

\section{Finite temperature correlation function}
Local spin DSF at finite temperature follows
\be
C^{\mathcal{O}}(\omega,T) = \int_{-\infty}^{\infty}dt \,e^{i\omega t}
\langle\mathcal{O}(t,0)\mathcal{O}^{\dagger}(0,0)\rangle_T,
\ee
where $\mathcal{O}$ denotes the field theory counterpart of local spin operators.
Using field theory language, the DSF can be expressed in the Lehmann spectral representation as,
\be
C^{\mathcal{O}}(\omega,T)= \frac{1}{\mathcal{Z}}\sum_{i,f}C^{\mathcal{O}}_{i,f}(\omega,T),
\label{eq:co}
\ee
with $C_{i,f}^{\mathcal{O}}$ labelling the contribution of excitation from $\#i$-particle states to $\#f$-particle states and
the partition function
$\mathcal{Z}=\text{Tr} e^{-H/T}=\sum_{n=0}^{\infty}\mathcal{Z}_n$.
Explicitly,
\be \ba
&C_{i,f}^{\mathcal{O}}(\omega,T) =
\sum_{\mathbf{i},\mathbf{f}}
\int\frac{d\theta_1^{\prime}... d\theta_{i}^{\prime}}{(2\pi)^{i} \mathcal{A}_{\mathbf{i}}}
\int\frac{d\theta_1... d\theta_{f}}{(2\pi)^{f} \mathcal{A}_{\mathbf{f}}}e^{-E_{\mathbf{i}}/T}
\\&\cdot
|\langle P_1^{\prime}(\theta_1^{\prime})... P_{i}^{\prime}(\theta_{i}^{\prime})|\mathcal{O}|P_1(\theta_1)... P_{f}(\theta_{f})\rangle|^2\delta(\omega+E_{\mathbf{i}}-E_{\mathbf{f}}),
\label{eq:general_Cmn}
\ea \ee
and
\be \ba
\mathcal{Z}_n &= \sumint_{\mathbf{n}}\frac{d\theta_1... d\theta_n}{(2\pi)^n\mathcal{A}_{\mathbf{n}}}e^{-\frac{E_{\mathbf{n}}}{T}}
\langle P_1(\theta_1)... P_n(\theta_n)|P_1(\theta_1)... P_n(\theta_n)\rangle,
\ea \ee
where the Boltzmann constant $k_B = 1$, $\mathbf{i}$ labels asymptotic state containing $i$ particles,
and $\mathcal{A}_{\mathbf{i}}=\prod_{l\in \mathbf{i}}n_l!$ with $n_l$ counting the particle number of type $l$.
At low temperature $ T\ll m_{B_1}$ with the Boltzmann factor $\exp(-E_{\mathbf{i}}/T)$ 
serving as a controlled parameter,
a regularized linked cluster expansion can be obtained \cite{Pozsgay_2010}
\be \ba
C^{\mathcal{O}}(\omega,T)=\sum_{i,f}^{\infty}D^{\mathcal{O}}_{i,f}(\omega,T),
\label{eq: cluster}
\ea \ee
where $D^{\mathcal{O}}_{0,f} = C^{\mathcal{O}}_{0,f}$, $D^{\mathcal{O}}_{1,f} = C^{\mathcal{O}}_{1,f}-\mathcal{Z}_1C^{\mathcal{O}}_{0,f-1}$,
$D^{\mathcal{O}}_{2,f} = C^{\mathcal{O}}_{2,f}-\mathcal{Z}_{1}C^{\mathcal{O}}_{1,f-1}+(\mathcal{Z}^2_1-\mathcal{Z}_2)C^{\mathcal{O}}_{0,f-2},
\; \cdots$.

\section{Non-vanishing $B_1\to B_1$ channel}
\begin{figure}[h]
    \centering
    \includegraphics[width=0.4\textwidth]{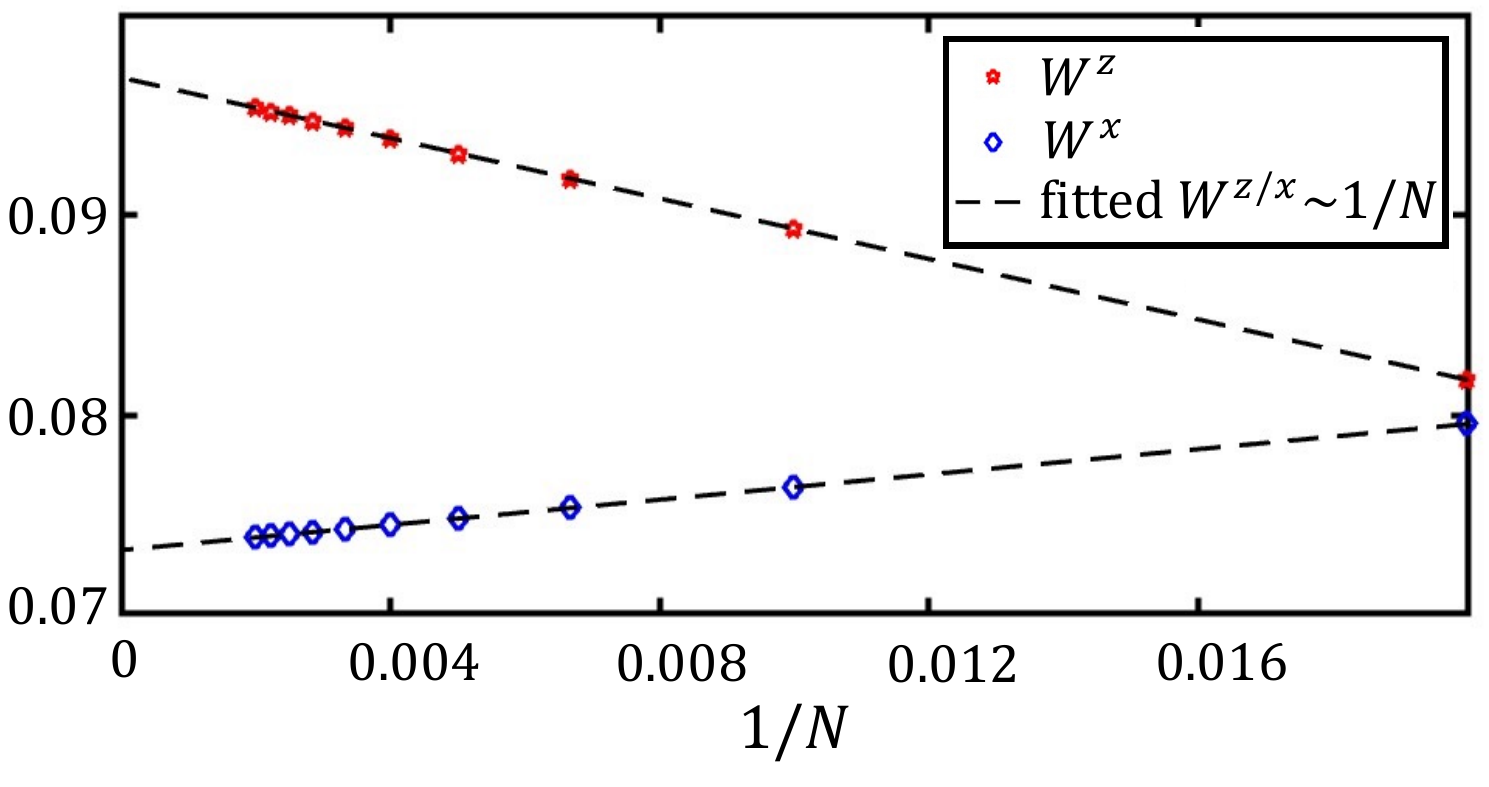}
    \caption{Relation between $W^{z/x}$ and $N$ obtained from DMRG calculation with $J_i=0.1$ and $J=1$.
    Dashed lines show the fitting function $W^z=-1.5172/N+0.0969$ and $W^x=-0.6331/N+0.0732$.}
    \label{fig:b1}
\end{figure}

Because the local spin along $y$ or $z$ direction is highly non-local
in the IIFT, $C^z$ and $C^y$ [Eq.~\eqref{eq:co}] are beyond
IIFT analytical form factor scheme.
However, their thermal behaviors can be determined numerically.
The first symmetry allowed channel in $C^z$ comes from $|B_1\rangle\to|B_1\rangle$,
dominant by zero momentum mode.
Its non-vanishing spectral weight can be confirmed by calculating $|\langle B_1(0)|\sigma^{(1,2)z}(0)|B_1(0)\rangle|^2$ in parallel to Eq.~(7),
effectively captured by $W^z$ defined through
\be
W^\alpha\equiv\sum_{n}[{}_\mathcal{L}\langle B_1(0)|\sigma^{\alpha(1,2)}_n|B_1(0)\rangle_{\mathcal{L}}]^2/N,
\ee
where $\alpha=x,\,z$, $N$ is the system size and $\mathcal{L}$ indicates the states for the lattice model.
$|B_1(0)\rangle_\mathcal{L}$ stands for the first excited state of the lattice Hamiltonian [Eq.~(1)]
Through DMRG calculation,
Fig.~\ref{fig:b1} shows that both $W^x$ and $W^z$ converge to a finite value as the lattice size $N\to \infty$,
suggesting the leading behavior $C^{z}\sim e^{-m_{B_1}/T}$.

\section{Relation between the gap and interchain coupling}
In the IIFT [Eq. (3)],
relation between the rescaled interchain coupling $\lambda$ and the gap $\Delta$ is known as \cite{BASEILHAC2001607}
\be
\lambda = -\frac{2\Gamma(1/8)}{\pi\Gamma(7/8)}\left[\frac{\Delta\sqrt{\pi}}{4\sin\frac{\pi}{14}}\frac{\Gamma(4/7)}{\Gamma(1/14)}\right]^{4/7}\equiv \mathcal{C}_1\Delta^{7/4}.
\ee
To obtain the the gap in terms of lattice model parameters,
a scaling relation is needed for the interchain coupling of discrete and continuum cases.
Since the model is obtained from two quantum 
critical transverse field Ising chains (TFICs) with weak interchain Ising coupling,
TFIC provide us with a starting point.
Following similar strategy in Ref. \cite{PhysRevB.110.195101},
to fix the normalization,
we have
\be
\langle\gamma\int dx\sigma^{(1)}(x)\sigma^{(2)}(x)\rangle=\langle JJ_i \sum_{j=1}^N \sigma_j^{z(1)}\sigma_j^{z(2)}\rangle.
\label{eq:norm}
\ee
Relation between $J_i$ and $\gamma$ should be recovered from the decoupled limit of Eq.~\eqref{eq:norm},
s.t. $\gamma L\langle\sigma\rangle^2 = JJ_iN\langle\sigma^z\rangle^2$.
Using the results in TFIC \cite{McCoyWu1973}
\be\ba
&\langle\overline{\sigma}\rangle^2 = \left(2^{1/12}e^{-1/8}\mathcal{A}^{3/2}M^{1/8}\right)^{2},\quad \mathcal{A}=1.2824271291...\\
&M=2J(1-h_x),\quad\langle\sigma^{z(1,2)}_j\rangle=\left(1-h_x^2\right)^{1/8},
\ea\ee
we substitute the expressions in Eq.~\eqref{eq:norm} and take the $h_x\to1$ limit,
obtaining
\be
\gamma = \frac{2J^{7/4}J_i}{2^{1/6}e^{-1/4}\mathcal{A}^{3}}.
\ee
Finally,
consider the normalization before and after bosonization that
\be
\langle\gamma\int dx\sigma^{(1)}(x)\sigma^{(2)}(x)\rangle=\langle\lambda\int dx\cos\phi(x)/2\rangle,
\ee
the coupling constants are related by $\gamma=\sqrt{2}\lambda$ as shown in \cite{BASEILHAC2001607}.
As a result,
\be
\Delta = \left(\frac{2\sqrt{2}}{\mathcal{C}_12^{1/6}e^{-1/4}\mathcal{A}^{3}}\right)^{4/7}J_i^{4/7}J.
\label{eq:scale}
\ee
By setting $J=1$,
Eq.~\eqref{eq:scale} implies that $\Delta=2.3797J_i^{4/7}$.
This is in good agreement with the numerical results shown in Fig.~\ref{fig:scale}.

\bibliography{Refs}
\end{document}